\documentclass[aoas,nameyear,seceqn,dvips]{arximspdf}
\usepackage{graphics}
\doi{10.1214/10-AOAS333}
\volume{4}
\issue{3}
\pubyear{2010}
\firstpage{1558}
\lastpage{1578}

\makeatletter
\newtheorem{theorem}{Theorem}
\newproclaim{eg}{Example}

\def\eqref#1{(\ref{#1})}

  \let\sv@tabnotetext\tabnotetext
  \let\sv@tabnotemark@fmt\tabnotemark@fmt
   \long\def\legend#1{{\let\tabnote@indent\leavevmode\sv@tabnotetext[]{}{#1}}}

\makeatother

\begin{document}
\begin{frontmatter}

\title{Accounting for choice of measurement scale in extreme value modeling}
\runtitle{Measurement scale in extreme value modeling}

\begin{aug}
\author[a]{\fnms{J. L.} \snm{Wadsworth}\corref{}\thanksref{t2}\ead[label=e1]{j.wadsworth@lancaster.ac.uk}},
\author[a]{\fnms{J. A.} \snm{Tawn}}
\and
\author[b]{\fnms{P.} \snm{Jonathan}}

\thankstext{t2}{Supported through a CASE studentship by the EPSRC and
Shell Research.}

\runauthor{J. L. Wadsworth, J. A. Tawn and P. Jonathan}

\affiliation{Lancaster University, Lancaster University and\break Shell
Technology Centre Thornton}

\address[a]{J. L. Wadsworth\\ J. A. Tawn\\ Department of Mathematics and Statistics\\ Lancaster
University\\ Lancaster LA1 4YF\\ United Kingdom\\ \printead{e1}}
\address[b]{P. Jonathan\\ Shell Technology Centre Thornton\\ P.O. Box CH1 3SH\\ Chester\\ United Kingdom}

\end{aug}

% HISTORY:
\received{\smonth{10} \syear{2009}}
\revised{\smonth{12} \syear{2009}}

% ABSTRACT
%
\begin{abstract}
We investigate the effect that the choice of measurement
scale has upon inference and extrapolation in extreme value analysis.
Separate analyses of variables from a single process on scales which are
linked by a nonlinear transformation may lead to discrepant conclusions
concerning the tail behavior of the process. We propose the use of a
Box--Cox power transformation incorporated as part of the inference
procedure to account parametrically for the uncertainty surrounding
the scale of extrapolation. This has the additional feature of
increasing the rate of convergence of the distribution tails to
an extreme value form in certain cases and thus reducing bias in
the model estimation. Inference without reparameterization is
practicably infeasible, so we explore a reparameterization which
exploits the asymptotic theory of normalizing constants required for
nondegenerate limit distributions. Inference is carried out in a Bayesian
setting, an advantage of this being the availability of posterior predictive
return levels. The methodology is illustrated on both simulated data and
significant wave height data from the North Sea.
\end{abstract}

% KEYWORDS
%

\end{frontmatter}

%s1 ###
\section{Introduction}
\label{sec:Introduction}

The usual objective of extreme value analysis is to use sample data
from rare events of a process to make rational predictions about the
likely levels of future extremes of the process. To do this, one models
extreme data using an asymptotically justified probability model. The
most fundamental such example is the generalized extreme value (GEV)
distribution. The GEV arises as the limiting law for appropriately
normalized maxima of independent and identically distributed random
variables, under weak conditions discussed in Section~\ref
{sec:TheoryAndMethods}; it is a three parameter distribution with
distribution function
\begin{eqnarray*}
G(x) &=& \exp\biggl\{-\biggl[1+\frac{\xi}{\sigma}(x-\mu)\biggr]^{-1/\xi}_+\biggr\},
\end{eqnarray*}
where $\mu, \sigma> 0, \xi$ are respectively location, scale and
shape parameters, and $z_+=\max\{0,z\}$. This distribution is herein
denoted $\operatorname{GEV}(\mu,\sigma,\xi)$. The cases $\xi>0$,
$\xi=0$
(interpreted as the limit $\xi\rightarrow0$) and $\xi<0$ are
sometimes referred to as the Fr\'{e}chet, Gumbel and Negative Weibull
types, respectively. Other approaches to modeling extreme data are
discussed in Section~\ref{sec:Methodology}. Mathematical details of
univariate extreme value theory for stationary processes are covered
extensively in \citet{LLR}, while more statistical aspects are
treated in, for example, \citet{Coles}.

There are many applications of extreme value analysis where data
pertaining to the same physical process may naturally be measured on
more than one scale. If the transformation between measurement scales
is linear, the appropriate type of extreme value distribution remains
unaltered. If, on the other hand, a nonlinear transformation is
applied, different limiting distributions may be appropriate. Applying
extreme value methods to the data on these different scales can lead to
disparate conclusions regarding future extremes. This paper proposes
methodology which allows the modeler to take into account their
uncertainty over the scale upon which to conduct extreme value
analysis.

As a motivating example consider the following. In ocean engineering,
significant wave height ($H_s$), defined as four times the standard
deviation of displacement from mean sea level, is a measure of ocean
energy. Understanding of the extremes of this variable is vital for
offshore structural design. However, one might equally wish to consider
the extremes of the drag force induced by the waves on a fixed offshore
structure, a variable which is proportional to the square of $H_s$
[\citet{Drag}]. Although the two variables are measurements of the same
physical process, differing conclusions may be derived concerning their
tail behavior. For the wave height data to be considered in
Section~\ref{sec:Examples}, a simple likelihood-based analysis of
weekly maxima of $H_s$ produces a 100-year return level estimate and
95\% confidence interval of 14.66 meters (13.63, 16.35). However,
analyzing $H_s^2$ instead, then back-transforming the results to the
$H_s$ scale, the estimate becomes 16.27 meters (14.51, 18.92).
Furthermore, the estimated shape parameters of the two variables differ
markedly: for $H_s$, $\hat{\xi} = -0.12$ ($-$0.17, $-$0.06), whereas for
$H_s^2$, $\hat{\xi}=0.11$ (0.04, 0.19). These results suggest
light-tailed behavior with a finite upper end point for $H_s$, yet
heavy-tailed behavior with no finite upper end point for $H_s^2$. Such
a situation gives rise to increasingly discrepant return level
inferences with lengthening return period. It seems natural therefore
to account for this uncertainty over the scale on which to extrapolate
as part of the inference.

We approach this problem by incorporating a power transformation into
the inference procedure; specifically, we use the well-known Box--Cox
transformation [\citet{BoxCox}]. This transformation offers the
possibility of improving the rate of convergence to the limiting
extreme value form, since different distributions converge at different
rates. This type of transformation restricts the methodology to cases
where the extreme data are strictly positive, however, this encompasses
a wide variety of practical problems. Use of the Box--Cox transform has
been previously considered by \citet{Teugels} as a way of improving
the rate of convergence, and in Section~\ref{sec:TheoryAndMethods} we
discuss how part of our work is related to theirs. However, their work
is purely probabilistic and, unlike ours, does not extend to consider
use of the theory as a statistical technique. The use of the Box--Cox
transformation in extreme value analysis has also been considered in an
entirely different context in the work of \citet{EastoeTawn}. In their
work the motivation was the standardization of nonstationary data prior
to the consideration of extreme values.

We choose to adopt a Bayesian methodology for our inferential
procedures, proceeding via Markov chain Monte Carlo (MCMC). The
Bayesian framework allows us to produce particularly useful posterior
summaries incorporating uncertainty from both the data and the
parameters. In particular, it enables the calculation of a posterior
predictive distribution, which provides a single useful summary of the
likelihood of future extremes under the two stated sources of
uncertainty [\citet{CT96}].

Moving from the usual three parameter extreme value models to four
parameter models including a Box--Cox parameter necessitates a
reparameterization. The theory we exploit to derive our
reparameterization is presented in Section~\ref{sec:TheoryAndMethods},
including a discussion on the rate of convergence. In Section~\ref
{sec:Methodology} we outline our reparameterizations and discuss
associated inference methods. In Section~\ref{sec:Examples} we
illustrate the methodology on simulated data and the aforementioned
significant wave height data. A discussion of the work and outstanding
issues is given in Section~\ref{sec:Discussion}.

%s2 ###
\section{Theory}
\label{sec:TheoryAndMethods}

%s2.1 ###
\subsection{Asymptotic and penultimate theory}
\label{sec:NormalizingConstants}

Suppose $X_1,\ldots,X_n$ are independent and identically distributed
according to a probability law with distribution function $F_X$, with
density $f_X$. In what follows it will be assumed that $F_X(x)$ is
twice differentiable for all sufficiently large $x$. Let $Y$ denote
these random variables after the application of a Box--Cox
transformation; that is, $Y = \{X^\lambda-1\}/\lambda$, $\lambda\in
\mathbb{R}$, the case $\lambda=0$ taken as $Y=\log X$, with
distribution function $F_Y$ and density $f_Y$. Define $M_{X,n} = \max\{
X_1,\ldots,X_n\}$. The extremal types theorem
[\citet{Fisher-Tippett}]
states that if there exist sequences of constants $\{a_{X,n}>0\}$, $\{
b_{X,n}\}$ such that as $n\rightarrow\infty$
%
%e2.1 ###
\begin{eqnarray}\label{gev}
\mathsf{P}\biggl(\frac{M_{X,n}-b_{X,n}}{a_{X,n}}\leq x\biggr) \stackrel
{w}{\longrightarrow} G(x)
\end{eqnarray}
for some nondegenerate limit distribution $G(x)$, then $G$ is
necessarily\vspace*{-1pt} of a generalized extreme value type. The symbol `$\stackrel
{w}{\rightarrow}$' denotes weak convergence of the distribution
functions.

Let $\{a_{X,n}\}$, $\{b_{X,n}\}$ henceforth specifically denote the
normalizing sequences which lead to a $\operatorname{GEV}(0,1,\xi_X)$ limit
distribution for the $M_{X,n}$. \citet{Smith87} shows that the
sequences $\{a_{X,n}\}$, $\{b_{X,n}\}$, and the shape parameter $\xi
_X$ can be found as follows. Let $h_X(x) = \{1-F_X(x)\}/f_X(x)$ denote
the reciprocal hazard function of the parent distribution $F_X$. Then
%
%e2.2 ###
\begin{eqnarray}\label{anxbnxxix}
\qquad b_{X,n} = F_X^{-1}(1-1/n),\qquad  a_{X,n} = h_X(b_{X,n}),\qquad  \xi_{X} = \lim
_{x \rightarrow x^F}h'_X(x)
\end{eqnarray}
with $x^F = \sup\{x\dvtx F_X(x)<1\}$, that is, the upper end point of
the distribution. A~finite value for $\xi_X$ given by limit~\eqref
{anxbnxxix} and our assumptions on $F_X$ are sufficient for weak
convergence~\eqref{gev} and necessary and sufficient for both
convergence of the densities and derivatives of the densities to those
of the limiting extreme value form [\citet{Pickands86}, Theorem~5.2] and
we assume this applies throughout.

The usual premise of extreme value modeling is to assume that the
limiting form~\eqref{gev} holds exactly for some finite $n$. \citet
{Fisher-Tippett} and \citet{Smith87} propose an approximation of
limit~\eqref{gev} by $M_{X,n} \stackrel{.}\sim\break\operatorname
{GEV}(b_{X,n}, a_{X,n},
\xi_{X,n})$ with $\xi_{X,n} = h'_X(b_{X,n})$, referred to as the
penultimate approximation to the shape parameter. From~\eqref
{anxbnxxix} we see $\xi_{X} = \lim_{n \rightarrow\infty}\xi
_{X,n}$. For inference purposes we therefore assume a three parameter
model $M_{X,n} \stackrel{.}\sim\operatorname{GEV}(\beta_X,\alpha
_X,\gamma_X)$,
where we reserve the notation $\xi_X$ for the limiting shape
parameter. The proposal of this paper is to generalize this modeling
assumption to
\begin{eqnarray*}
M_{Y,n} = \frac{M_{X,n}^\lambda-1}{\lambda} \stackrel{.}\sim
\operatorname{GEV}
(\beta_Y,\alpha_Y,\gamma_Y),
\end{eqnarray*}
thereby incorporating a form of parametric scale uncertainty into the
inference procedure. This gives a four parameter extreme value model,
with canonical parameterization $\{\beta_Y,\alpha_Y,\gamma_Y,\lambda
\}$. The complex nature of the relationships between these parameters,
however, makes direct inference practicably infeasible (see Figure~\ref
{fig:Ex1Post} in Section~\ref{sec:Examples} for an illustration).
Thus, a reparameterization to obtain more orthogonal relationships is
necessary. Our strategy for orthogonalization relies upon obtaining $\{
a_{Y,n}\}, \{b_{Y,n}\}, \xi_{Y,n}$ in terms of the associated
quantities for the original~$X$ variables.

\begin{theorem}
\label{th1}
Under the conditions such that convergence~\eqref{gev} holds, with
norming sequences $\{a_{X,n}\}, \{b_{X,n}\}$ producing the
$\operatorname{GEV}
(0,1,\xi_X)$ limit, then
\begin{eqnarray*}
\mathsf{P}\biggl(\frac{M_{Y,n}-b_{Y,n}}{a_{Y,n}}\leq y\biggr) \stackrel
{w}{\longrightarrow} G_Y(y) = \exp\{-[1+\xi_Y y]^{-1/\xi_Y}_+\}
\end{eqnarray*}
holds for some finite $\xi_Y$ when
%
%e2.3 ###
\begin{eqnarray}\label{bnyany}
b_{Y,n} = \frac{(b_{X,n})^\lambda-1}{\lambda},\qquad   a_{Y,n} =
a_{X,n}(b_{X,n})^{\lambda-1}.
\end{eqnarray}
Furthermore, if $F_X$ is twice differentiable for sufficiently large
$x$, then the limiting shape parameter $\xi_Y$ takes the form
%
%e2.4 ###
\begin{eqnarray}\label{xiy1}
\xi_Y = \xi_X +\lim_{x\rightarrow x^F}\frac{h_X(x)}{x} (\lambda
-1)
\end{eqnarray}
with the penultimate approximation to this being given by
%
%e2.5 ###
\begin{eqnarray}\label{xiyn}
\xi_{Y,n} = \xi_{X,n} + \frac{a_{X,n}}{b_{X,n}}(\lambda-1).
\end{eqnarray}
For any such distribution which has $\xi_X \leq0,$ then $\xi_Y = \xi_X$.
\end{theorem}

See the \hyperref[sec:Appendix]{Appendix} for a proof. Equations~\eqref{bnyany} and~\eqref
{xiy1} are used in Section~\ref{sec:Methodology} to motivate
reparameterizations for the statistical models for block maxima and
threshold exceedances. Note that when $F_X$ is in the domain of
attraction of a Negative Weibull or Gumbel limit, then $F_Y$ is in the
same domain of attraction; only those distributions which have a Fr\'
{e}chet limit can be coerced into a different domain. However, as we
are never practically in the limit, and $h_X(x)/x>0$ for $x>0$, values
of $\lambda$ other than 1 will alter the penultimate approximation and
thus change our practical estimation of the shape parameter for the
transformed variables regardless of domain of attraction.

%s2.2 ###
\subsection{Rate of convergence}
\label{sec:RateOfConvergence}

It was noted in Section~\ref{sec:Introduction} that the rate of
convergence to the limiting extreme value distribution may be altered
by a power transformation. In \citet{Teugels} the theory of regular
variation is exploited to show what the optimal values of the power
transformation parameter should be to maximize the rate of convergence
in the case where $\xi_X \geq0$. Under our assumptions on $F_X$, we
derive similar limiting results to \citet{Teugels} for $\xi_X\geq0$,
but also consider the case $\xi_X <0$ and the penultimate
approximations. In particular, the examples studied in \citet{Teugels}
satisfy our assumptions.

We use approximations developed by \citet{Smith87} as a basis for
discussion on rates of convergence. Smith shows that for $h'_X \not
\equiv0$ one may write
%
%e2.6 ###
\begin{eqnarray}\label{approx}
\{F_X (a_{X,n}x+b_{X,n})\}^n = \exp\bigl\{-[1+h'_X(z)x]^{-1/h'_X(z)}\bigr\} +
O(n^{-1})
\end{eqnarray}
for some $z\in[\min\{a_{X,n} x + b_{X,n}, b_{X,n}\}, \max\{a_{X,n} x
+ b_{X,n}, b_{X,n}\}]$. For $h'_X \equiv0$ the first term on the RHS
is $e^{-x}$. It follows that the rate of pointwise distributional
convergence is
\[
\max\bigl\{O\bigl(|h_X'(b_{X,n})-\xi
_X|\bigr),O\bigl(|h_X'(a_{X,n}x+b_{X,n})-h_X'(b_{X,n})|\bigr), O(n^{-1})\bigr\}.
\]
We focus on demonstrating how an improved rate of convergence is
possible when this rate is equal to $O(|h_X'(b_{X,n})-\xi_X|)$. This
will in fact be the case if $O(\{h_X(b_{X,n})\}^r
h_X^{(r+1)}(b_{X,n}))\leq O(|h_X'(b_{X,n})-\xi|), \forall r\geq1$.
This is a condition satisfied by a wide range of theoretical examples,
including Examples~\ref{eg1}--\ref{eg4} below. For such distributions
an improved rate of convergence will be achieved if
$O(|h_Y'(b_{Y,n})-\xi_Y|) < O(|h_X'(b_{X,n})-\xi_X|)$. By
expressions~\eqref{anxbnxxix}, \eqref{xiy1} and~\eqref{xiyn},
%
%e2.7 ###
\begin{eqnarray}\label{rate1}
\qquad &&|h_Y'(b_{Y,n})-\xi_Y|\nonumber
\\[-8pt]\\[-8pt]
&&\qquad = \biggl|h'_X(b_{X,n}) +\frac
{h_X(b_{X,n})}{b_{X,n}}(\lambda-1) - \lim_{x\rightarrow x^F} \biggl\{
h'_X(x) +\frac{h_X(x)}{x}(\lambda-1)\biggr\}\biggr|\nonumber
\\\label{rate2}
&&\qquad = |h'_X(b_{X,n})-\xi_X |\biggl|1+(\lambda-1)\frac{
{h_X(b_{X,n})/b_{X,n}}-\lim_{x\rightarrow x^F}
{h_X(x)/x}}{h'_X(b_{X,n})-\xi_X}\biggr|.
\end{eqnarray}
Equation~\eqref{rate2} demonstrates accelerated convergence under the
transformation if the second term on the RHS improves the order. This
is the case for any value of $\lambda$ which gives convergence of this
second term to 0. In particular, this means there is a sequence of
$\lambda$ values, denoted $\{\lambda_n^*\}$ and given by
%
%e2.8 ###
\begin{eqnarray}\label{lam}
\lambda^*_n \sim1-\frac{h'_X(b_{X,n})-\xi_X}{
{h_X(b_{X,n})/b_{X,n}}-\lim_{x\rightarrow x^F}{h_X(x)/x}},
\end{eqnarray}
which provide the best rate of convergence under any such
transformation.

For statistical applications the convergence rate of densities is also
relevant. Pointwise
density convergence entails an additional error term of
$O(|h_X(a_{X,n}x+b_{X,n})/h_X(b_{X,n})
- (1+\xi_X x)|)$. The conditions on $h_X$ which allow us to consider
$O(|h_X'(b_{X,n})-\xi_X|)$
for distribution function convergence give that
\begin{eqnarray*}
O\bigl(|h_X(a_{X,n}x+b_{X,n})/h_X(b_{X,n}) - (1+\xi_X x)|\bigr) = O\bigl(|h_X'(b_{X,n})-\xi_X|\bigr).
\end{eqnarray*}
For each of our variety of examples below pointwise density convergence
occurs at the same rate as distribution function convergence, and any
$\lambda$ which improves the rate of distribution function convergence
also improves that of the density function. We can of course never
check any of these conditions in practice, and it is our data rather
than any theoretical knowledge which point to a value of $\lambda$; as
such, we presume that by pursuing this approach we at least do not lose
in terms of convergence rate of the densities.

Below we provide illustrations for four different classes of
distribution, largely following the examples laid out in \citet
{Smith87}. We make the corresponding assumptions that the relationships
in Examples~\ref{eg1}--\ref{eg3} are twice-differentiable, in the
sense that we can differentiate term-wise without affecting the
$O$-term representation. Table~\ref{tab:xi} summarizes the shape
parameters for these examples, alongside the order of convergence of
the penultimate approximations. Also detailed are values of $\lambda$,
denoted $\lambda^*$, which provide an improved rate of convergence.
Note that these values are the limiting values of the sequence $\{
\lambda_n^*\}$, where such a limit exists.

\begin{eg}\label{eg1}
$x^F=+\infty$; $ \alpha,\beta,\varepsilon,C>0; D\in\mathbb{R}$,
\begin{eqnarray*}
1-F_X(x) = Cx^{-\alpha}\{1+Dx^{-\beta}+O(x^{-\beta-\varepsilon})\}
 \qquad \mbox{as $x\rightarrow x^F$}.
\end{eqnarray*}
This class belongs to the Fr\'{e}chet domain of attraction. Examples
include the Pareto, $t$, $F$ and Cauchy distributions. If $D\neq0$, then
taking $\lambda^* = \beta$ forces the leading term in $|\xi
_{Y,n}-\xi_Y|$ to vanish, thus improving the convergence rate.
\end{eg}

\begin{eg}\label{eg2}
$x^F<+\infty$; $ \alpha,\beta,\varepsilon,C>0; D\in\mathbb{R}$,
%
%e2.9 ###
\begin{eqnarray}
1-F_X(x) = C(x^F-x)^{\alpha}\bigl\{1+D(x^F-x)^{\beta}+O\bigl((x^F-x)^{\beta
+\varepsilon}\bigr )\bigr\}\nonumber
\\
 \eqntext{\mbox{as $x\rightarrow x^F$}.}
\end{eqnarray}
This class belongs to the Negative Weibull domain of attraction.
Examples are distributions with bounded upper tails, such as the beta,
along with various truncated distributions. Depending on the value of
$\beta$, the best rate of convergence is either given by $\lambda
^*=1$ ($\beta>1$), or if $\beta<1$, the value of $\lambda$ is
asymptotically inconsequential, and in this case the sequence $\{
\lambda^*_n\}$ has no limit.
\end{eg}

\begin{eg}\label{eg3}
$x^F=+\infty$; $ \alpha>-1; \varepsilon>0; C>0$,\vspace*{-1pt}
\begin{eqnarray*}
h_X(x)=\frac{1-F_X(x)}{f_X(x)} = Cx^{-\alpha}\{1+O(x^{-\varepsilon
})\}
\qquad \mbox{as }x\rightarrow x^F.%& \mbox{ as $x\rightarrow x^F$}.
\end{eqnarray*}
This class belongs to the Gumbel domain of attraction. Examples include
exponential ($\alpha=0$), normal ($\alpha=1$), Weibull ($\alpha
=\gamma-1$, for Weibull shape parameter~$\gamma$) and gamma ($\alpha
=0$). Taking $\lambda^*=\alpha+1$ improves the rate of convergence,
again via elimination of the leading order term in $|\xi_{Y,n}-\xi
_Y|$.

%t1 ###
\begin{table}[t]
\tabcolsep=0pt
\caption{Shape parameters and the leading order terms from the
penultimate approximations for Examples~\textup{\protect\ref{eg1}--\protect\ref{eg4}}. Three
subcases for Example~\textup{\protect\ref{eg4}}: \textup{(i)} $\gamma<0$, \textup{(ii) }$\gamma=0$,
\textup{(iii)} $\gamma>0$}\label{tab:xi}
\begin{tabular*}{\tablewidth}{@{\extracolsep{4in minus 4in}}lccccc@{}}
\hline
\textbf{Example}& $\bolds{\xi_X}$ & $\bolds{\xi_{X,n}-\xi_X}$ & $\bolds{\xi_Y}$ & $\bolds{\xi_{Y,n}-\xi_Y}$
&$\bolds{\lambda^*}$
\\
\hline
\ref{eg1} & $1/\alpha$ & \multicolumn{1}{l}{$ \sim\frac{D\beta(\beta-1)}{\alpha
^2}(nC)^{-\beta/\alpha}$ }& $\lambda/\alpha$&  \multicolumn{1}{l}{$\sim\frac{D\beta
(\beta-\lambda)}{\alpha^2}(nC)^{-\beta/\alpha}$} & $\beta$
\\[3pt]
\ref{eg2} & $-1/\alpha$ &  \multicolumn{1}{l}{$\sim\frac{D\beta(\beta+1)}{\alpha
^2}(nC)^{-\beta/\alpha}$} & $-1/\alpha$&  \multicolumn{1}{l}{$\sim\frac{D\beta(\beta
+1)}{\alpha^2}(nC)^{-\beta/\alpha}-\frac{\lambda-1}{x^F\alpha}
(nC)^{-1/\alpha}$} &1
\\[3pt]
\ref{eg3} & $0$ &  \multicolumn{1}{l}{$\sim-\alpha C b_{X,n}^{-\alpha-1}$} & $0$&  \multicolumn{1}{l}{$\sim
C(\lambda-(\alpha+1))b_{X,n}^{-\alpha-1}$} & $\alpha+1$
\\[2pt]
\ref{eg4}(i) & $\gamma$ & $\beta(n ^\gamma)$ & $\gamma$ & $\lambda
\beta(n^\gamma)$ &$0$\\
\ref{eg4}(ii) & $\beta$ &$0$& $\lambda\beta$ & $0$& \mbox{None}
\\
\ref{eg4}(iii) & \mbox{N/A}& \mbox{N/A}&$\gamma^\dagger
$&$0^\dagger$& $0$\\
\hline
\end{tabular*}
\legend{$^\dagger$If and only if $\lambda=0$.}\vspace*{-1pt}
\end{table}

In particular, note that for the normal distribution $\lambda^* = 2$
leads to faster convergence, the rate being improved from $O((\log
n)^{-1})$ to $O((\log n)^{-2})$. More generally for sub-asymptotic
levels, when~\eqref{lam} is used to obtain the appropriate sequence,
$\lambda^*_n\nearrow2$ as $n\rightarrow\infty$. This example is
revisited in Section~\ref{sec:NormalDistribution}.
\end{eg}

\begin{eg}\label{eg4}
$x^F=+\infty$ if $\gamma\geq0$, otherwise $x^F = e^{u-\beta/\gamma
}$; $\beta>0; \gamma,u\in\mathbb{R}$,\vspace*{-1pt}
\begin{eqnarray*}
1-F_X(x) = \biggl[1+\frac{\gamma}{\beta}(\log x-u)\biggr]_+^{-1/\gamma}.
\end{eqnarray*}
This corresponds to the class of log-Pareto distributions [\citet
{CormannReiss}]. For this class $\lim_{x\rightarrow x^F}h_X'(x)$ does
not exist if $\gamma> 0$; in this case the distribution is considered
`super-heavy-tailed' and falls into the domain of attraction of an
extreme value distribution if and only if the Box--Cox parameter
$\lambda=0$. This provides the most well-known example of a
distribution function outside any domain of attraction: $1-F_X(x) =
1/\log(x), x>e$, when $\gamma=\beta=u=1$.

When $\lim_{x\rightarrow x^F}h_X'(x)$ does not exist,~\eqref{rate2}
and~\eqref{lam} lack meaning, and one may revert to~\eqref{rate1} to
investigate whether any value of $\lambda$ which forces the existence
of $\lim_{y\rightarrow y^F}h_Y'(y)$ can be found. Direct consideration
of $h_Y'(y)$, in this case writing $x$ in place of $b_{X,n}$, yields
\begin{eqnarray*}
\beta+\gamma(\log x-u)+\gamma- (\lambda-1)\lim_{x\rightarrow
x^F}\{ \beta+\gamma(\log x-u) \},
\end{eqnarray*}
the limit of which can be seen to exist if and only if $\lambda=0$.
\end{eg}

%s3 ###
\section{Methodology}
\label{sec:Methodology}

%s3.1 ###
\subsection{Models}
\label{sec:Models}

The modeling setup we introduce for block maxima is
\begin{eqnarray*}
M_{X,n} \stackrel{.}\sim\operatorname{GEV}(\beta_X,\alpha_X,\gamma
_X), \qquad M_{Y,n}
= \frac{M_{X,n}^\lambda-1}{\lambda} \stackrel{.}\sim\operatorname
{GEV}(\beta
_Y,\alpha_Y,\gamma_Y).
\end{eqnarray*}
This provides parameter sets $\bolds{\theta}_{{X}} = \{\beta
_X,\alpha_X,\gamma_X\}$ and $\bolds{\theta}_{{Y}} = \{\beta
_Y,\alpha_Y,\gamma_Y, \lambda\}$. In particular, the shape parameter
$\gamma_Y$ is our finite sample approximation to $\xi_{Y,n}$, the
penultimate approximation to the limiting shape parameter $\xi_Y$.
Estimation of the parameter set $\bolds{\theta}_{{Y}}$ directly is
unwieldy. This is caused by the strong dependence introduced through
the additional parameter $\lambda$, as exhibited in the norming
constant and penultimate approximation expressions of equations~\eqref
{bnyany} and~\eqref{xiy1}; see also Figure~\ref{fig:Ex1Post} in
Section~\ref{sec:SimulatedDataExample}. Our approach to reducing the
dependence among the parameter set is described in the following
section.

The above description pertains specifically to the GEV model, however,
a common alternative to the block maxima approach in extreme value
analysis is to model all data which exceed some high threshold. The two
modeling strategies employed for this purpose are (i) model exceedances
via the generalized Pareto distribution [\citet{DavisonSmith90}], or
(ii) model exceedances using a non-homogeneous Poisson process [\citet
{Pickands71}]. Case (i) is essentially a reformulation of case (ii), so
we discuss here only the latter approach. The formal asymptotic
justification for the Poisson process model is that if we have a
sequence of two-dimensional point processes
\begin{eqnarray*}
P_n = \biggl\{\biggl(\frac{X_i-b_{X,n}}{a_{X,n}},\frac{i}{n+1}\biggr)\dvtx i=1,\ldots,n\biggr\},
\end{eqnarray*}
then on $(x_F^*,\infty)\times(0,1)$, where $x_F^* = \lim
_{n\rightarrow\infty}\{x_F-b_{X,n}\}/a_{X,n}$ with $x_F = \inf\{
x\dvtx F_X(x)>0\}$, $P_n\rightarrow P$, a Poisson process with intensity measure
\begin{eqnarray*}
\Lambda\{(x,\infty)\times(a,b)\} = (b-a)(1+\xi_X x)_+^{-1/\xi_X},
\qquad 0\leq a < b \leq1, x_F^*<x<\infty.
\end{eqnarray*}
The normalizing constants $\{a_{X,n}\}, \{b_{X,n}\}$ and the shape
parameter $\xi_X$ are exactly as before, thus, for statistical
inference on un-normalized data we model using a three parameter
nonhomogeneous Poisson process, denoted $\operatorname{PP}(\beta
_X,\alpha_X,\gamma
_X)$, with intensity measure
%
%e3.1 ###
\begin{eqnarray}\label{pp1}
\Lambda\{(x,\infty)\times(a,b)\} = (a-b)\biggl[1+\frac{\gamma_X}{\alpha
_X} (x-\beta_X)\biggr]_+^{-1/\gamma_X}.
\end{eqnarray}
This parameterization is easily unified with that of the GEV. If
observed data correspond to a particular number of blocks $N_B$, then
to estimate the GEV parameters corresponding to these block maxima, $\{
\beta_X', \alpha_X', \gamma_X'\}$, one assumes $N_B$ independent
replications of the Poisson process with $a=0$, $b=1$. Thus, the
statistical model becomes a Poisson process with intensity measure
%
%e3.2 ###
\begin{eqnarray}\label{pp2}
N_B \biggl[1+\frac{\gamma'_X}{\alpha'_X} (x-\beta'_X)\biggr]_+^{-1/\gamma
'_X}.
\end{eqnarray}
The relation between the parameters in~\eqref{pp1} with $a=0, b=1$
and~\eqref{pp2} are given by\vspace*{-2pt}
%
%e3.3 ###
\begin{equation}\label{pp3}
\qquad\ \hspace*{1,4pt} \gamma'_X=\gamma_X=\gamma,\qquad\hspace*{-0,5pt} \beta'_X = \beta_X-\dfrac{\alpha
_X}{\gamma}\biggl(1-\biggl(\dfrac{1}{N_B}\biggr)^{\gamma}\biggr),
\qquad\hspace*{-0,5pt}
  \alpha'_X=\alpha
_X\biggl(\dfrac{1}{N_B}\biggr)^{\gamma}.
\end{equation}
Both GEV and point process methods are considered in our examples in
Section~\ref{sec:Examples}, where the key focus of our modeling is
return level inference.

%s3.2 ###
\subsection{Reparameterization}
\label{sec:Reparameterization}

When fitting a $\operatorname{GEV}(\beta_Y,\alpha_Y,\gamma_Y)$
distribution to
$M_{Y,n}$, the parameters $\{\beta_Y, \alpha_Y\}$ are the unknown
quantities $\{b_{Y,n}, a_{Y,n}\}$. This is a direct consequence of
$a_{Y,n}, b_{Y,n}$ being specifically the sequences which give a
$\operatorname{GEV}
(0,1,\xi_Y)$ limit distribution for $M_{Y,n}$. Therefore, Theorem~\ref
{th1} leads to the reparameterizations
%
%e3.4 ###
\begin{eqnarray}\label{muysigy}
\beta_Y = \frac{\beta_X^\lambda-1}{\lambda},\qquad
\log\alpha_Y = (\lambda-1)\log\beta_X +\log\alpha_X,
\end{eqnarray}
the log function being used in the latter to both ensure the positivity
constraint is respected and to linearize dependence.
For $\gamma_Y$ the situation is slightly more subtle. Equation~\eqref
{xiyn} suggests taking
%
%e3.5 ###
\begin{eqnarray}\label{xiy}
\gamma_{Y} = \gamma_X + \frac{\alpha_X}{\beta_X}(\lambda
-1).
\end{eqnarray}
However, recalling equation~\eqref{approx}, one can see that the
estimable value of the shape parameter will not in general be
$h'_Y(b_{Y,n})$, but rather closer to being $h'_Y(b_{Y,n}+\varepsilon
)$, for some unknown $\varepsilon$. Thus, the parametric form~\eqref
{xiy} which is motivated by equation~\eqref{xiyn} is not strictly
appropriate, and the discrepancy between $b_{Y,n}$ and $b_{Y,n} +
\varepsilon$ can be sufficiently large that the structure~\eqref{xiy}
is a poor choice. This presents a problem finding a satisfactory
theoretical solution to the ratio in expression~\eqref{xiy} which
multiplies $\lambda-1$.

To overcome this, we have adopted the pragmatic solution of setting
%
%e3.6 ###
\begin{eqnarray}\label{xiy2}
\gamma_{Y} = \gamma_X + c(\lambda-1),
\end{eqnarray}
where $c$ is a fixed value estimated prior to inference. As
equation~\eqref{xiy2} corresponds to a linear relationship between
$\lambda$ and $\gamma_Y$, we used the shape of the profile likelihood
for $\{\gamma_Y,\lambda\}$ to identify the gradient of the
relationship. We estimate $c$ via calculating the profile
(log-)likelihood, $\mathrm{P} \ell(\gamma_Y,\lambda)$ on a fine
grid and
performing a weighted least squares fit to the grid points in order to
extract this slope. The weights are chosen at $\{\gamma_Y,\lambda\}$
to be $\exp[-2\{\mathrm{P} \ell(\hat{\gamma}_Y,\hat{\lambda
})-\mathrm{P} \ell(\gamma
_Y,\lambda)\}]$, thus ensuring that the ridge of high likelihood
dominates the fit and reduces sensitivity of the resulting estimate to
the choice of grid. Note that the calculation of $\mathrm{P} \ell
(\gamma
_Y,\lambda)$ over a particular region of interest presents no
difficulties, but full inference from the likelihoods
 for $\bolds{\theta}_{{Y}}$ is infeasible. This two-step approach to the
reparameterization has proven to work well in practice.

%s3.3 ###
\subsection{Inference}
\label{sec:Inference}

The likelihood functions for a general $\operatorname{GEV}(\beta
,\alpha,\gamma)$
distribution and $\operatorname{PP}(\beta,\alpha,\gamma)$ above a
threshold $u$
are given for $m$ independent and identically distributed data points by
\begin{eqnarray}\label{GEVlik}
\quad &&L_{\mathrm{GEV}}(\beta,\alpha,\gamma)\nonumber
\\[-8pt]\\[-8pt]
&&\qquad =\prod_{i=1}^m\exp\biggl\{
-\biggl[1+\frac{\gamma}{\alpha}(x_i-\beta)\biggr]_+^{-1/\gamma}\biggr\}\frac
{1}{\alpha}\biggl[1+\frac{\gamma}{\alpha}(x_i-\beta)\biggr]_+^{-1/\gamma-1},\nonumber
\\\label{PPlik}
&&L_{\mathrm{{PP}}}(\beta,\alpha,\gamma)\nonumber
\\[-8pt]\\[-8pt]
&&\qquad =\exp\biggl\{-N_{\mathrm{{B}}}\biggl[1+\frac{\gamma}{\alpha}(u-\beta)\biggr]_+^{-1/\gamma}\biggr\}
\prod_{i=1}^m\frac{1}{\alpha}\biggl[1+\frac{\gamma}{\alpha}(x_i-\beta
)\biggr]_+^{-1/\gamma-1},\nonumber
\end{eqnarray}
respectively, with $\{x_i\}$ representing realized block maxima and
threshold exceedances in equations~\eqref{GEVlik} and~\eqref{PPlik}
respectively. To extend these likelihoods to the 4 parameter case
simply requires that $u, x_i$ are replaced by $\{u^\lambda-1\}/\lambda
, \{x_i^\lambda-1\}/\lambda$, and that each term in the product is
multiplied by the Jacobian $x_i^{\lambda-1}$. In what follows,
reference to a `3 parameter model' relates directly to traditional
extreme value models whose likelihoods are given by equations~\eqref
{GEVlik} and~\eqref{PPlik}. Reference to a `4 parameter model'
pertains to our extension.

Equations~\eqref{muysigy} and~\eqref{xiy2} represent our
reparameterizations of $\bolds{\theta}_{{Y}}$ in terms of a new set
of parameters $\{\beta_X,\log\alpha_X,\gamma_X,\lambda\}$. As the
first three link clearly to inference for $M_{X,n}$, this allows
selection of good choices for parameter starting values by commencing
initially with a 3 parameter fit. In our algorithms vague Gaussian
priors (variance 10,000, centered on the estimates from the 3 parameter
fit) and Gaussian random walk sampling are used for $\beta_X, \log
\alpha_X, \gamma_X$, and a uniform prior with independent sampling
for $\lambda$. The parameter range for $\lambda$ is informed by
inspection of the profile likelihood $\mathrm{P} \ell(\gamma
_Y,\lambda)$.

The algorithm includes the constraint that if $\lambda<0$, $\gamma
_Y<0$, since the former implies a finite upper end point to the
distribution, which is only the case when the latter also holds.
Furthermore, in the case $\lambda<0$ this upper end point is $\{
(x^F)^\lambda-1\}/\lambda\leq-1/\lambda$, thus, we also impose the
constraint that the upper end point of the fitted GEV is $\beta
_Y-\alpha_Y/\gamma_Y\leq-1/\lambda$.

It was found that setting $N_{\mathrm{{B}}} \approx m$, the
number of threshold exceedances, in equation~\eqref{PPlik} improved
the mixing properties of the chain. This presents no major difficulties
since the equations in~\eqref{pp3} demonstrate how parameters
corresponding to different numbers of assumed blocks are linked. A
reason for improved mixing under this adjustment is that for $n$ total
observations and $m$ exceedances the location parameter $\beta$
becomes the $1-m/n$ quantile of the true underlying distribution $F_X$.
However, our fixed and known threshold $u$ is the empirical estimate of
this quantile, hence, this choice orthogonalizes the relationship
between $\beta$ and the other parameters.

The output of the MCMC leads to inference on return levels through
posterior distributions on specific quantiles, and via the posterior
predictive distribution. The $1/p$ block return level, $x_{1/p}$, which
is the $1-p$ quantile of the distribution is found via
%
%e3.7 ###
\begin{eqnarray}\label{xp}
x_{1/p} = [\lambda y_{1/p}+1]^{1/\lambda} ,
\end{eqnarray}
where
\begin{eqnarray*}
y_{1/p} = \beta_Y - \frac{\alpha_Y}{\gamma_Y}[1-\{-\log(1-p)\}
^{-\gamma_Y}].
\end{eqnarray*}
The $1/p$ block posterior predictive return level, denoted $\hat
{x}_{1/p}$, which corresponds to the $1-p$ quantile of the posterior
predictive distribution for $M_{X,n}$, is found by numerically solving
\begin{eqnarray*}
\mathsf{P}(M_{X,n}\leq\hat{x}_{1/p}|\mathbf{x}) = \int\mathsf{P}
(M_{Y,n}\leq\{\hat{x}_{1/p}^\lambda-1\}/\lambda|\bolds{\theta
}_{{Y}})p(\bolds{\theta}_{{Y}}|\mathbf{x})\,{d}\bolds{\theta}_{{Y}}
= 1-p,
\end{eqnarray*}
where $\mathbf{x}$ represents the realized data, either in block
maxima or threshold excess form. In practice, this is approximated
through a discrete integral over the MCMC output for
$\bolds{\theta}_{{Y}}$.

%s4 ###
\section{Examples}
\label{sec:Examples}

%s4.1 ###
\subsection{Simulated data examples}
\label{sec:SimulatedDataExample}

Two examples are presented. The first illustrates behavior when an
exact extreme value distribution is recoverable through a power
transformation. The second presents the case of the normal
distribution, demonstrating the practical effect of the differing rates
of convergence for transformed and untransformed variables. For each
example the burn-in period was 1000 iterations, with our reported
analyses based on the subsequent 10,000 draws.

%s4.1.1 ###
\subsubsection{Pre-transformed extreme value model}
\label{sec:ExtremeValueModel}

Data were simulated from a nonhomogeneous Poisson process with
parameters $\{\beta,\alpha,\gamma\} = \{15, 1.5, -0.25\}$ and the
threshold $u$ was fixed by the parameters so that $\Lambda\{(u,\infty
)\times(0,1)\} = 100{,}000$. The data were generated on the basis of
1000 blocks, {that is,} taking $N_{\mathrm{{B}}}$ in~\eqref
{PPlik} to be 1000. Three sub-samples of these data were analyzed:
\begin{enumerate}[1.]
\item Block maxima: 1000 maxima taken of blocks of length 100. These
data are exactly $\operatorname{GEV}(15,1.5,-0.25)$ distributed.

\item Largest 1000 data: threshold selected to retain only the largest
1000 points. Owing to the threshold stability property of the Poisson
process, these still have a $\operatorname{PP}(15,1.5,-0.25)$
distribution.

\item All data exceeding the smallest block maximum: threshold selected
to be equal to the minimum data point in data set~1. This
gave 6847 data points. Again these are $\operatorname{PP}(15,1.5,-0.25)$
distributed.
\end{enumerate}

As a testing ground for the ability of the methodology to detect a
`true' value of $\lambda$ when one exists, a square transformation was
pre-applied to data sets~1, 2 and~3, thus, they no longer followed the exact extreme value
distributions from which they were generated; these distributions being
recoverable, up to location and scale shifts, by taking $\lambda
=0.5$.

\begin{figure}[t]

\includegraphics{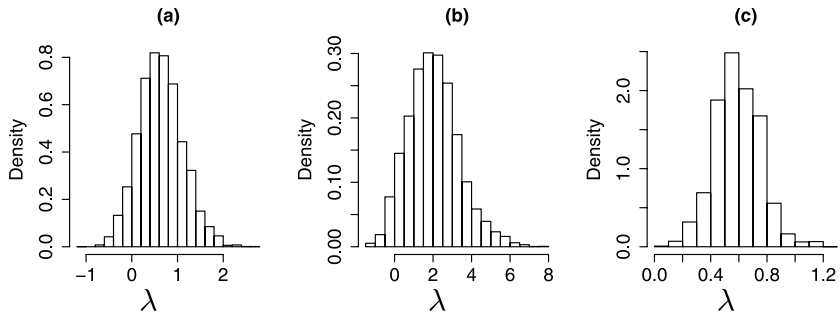}

\caption{Transformed extreme value model example: posteriors
for $\lambda$ from \textup{(a)} data set~1, prior range $[-2,3]$;
\textup{(b)} data set~2, prior range $[-2,8]$; \textup{(c)} data set~3, prior range $[0,2]$.}
\label{fig:PPLambdaPost}
\end{figure}

Figure~\ref{fig:PPLambdaPost} displays the posterior distributions for
$\lambda$ in each of the three scenarios. The ranges of the uniform
priors for $\lambda$ are detailed in the caption. Modes around
$\lambda=0.5$ are detectable in (a) and (c) (data sets~1
and~3), with the latter being much the more concentrated
density. The least information on $\lambda$ is obtained from data
set~2. This is explained by the relative extremity of the
data. The more extreme the data, the more the standard asymptotic
convergence arguments apply. That is, with data set~2, in
particular, the process is approximately Poisson regardless of the
transformation since we are still considering the largest $1\%$ of a
sample which is in the domain of attraction of an extreme value
distribution. Data set~3 contains a larger amount of
data, with the additional data being less extreme than that of data
set~2, thus producing the most informative posterior.

\begin{figure}[t]

\includegraphics{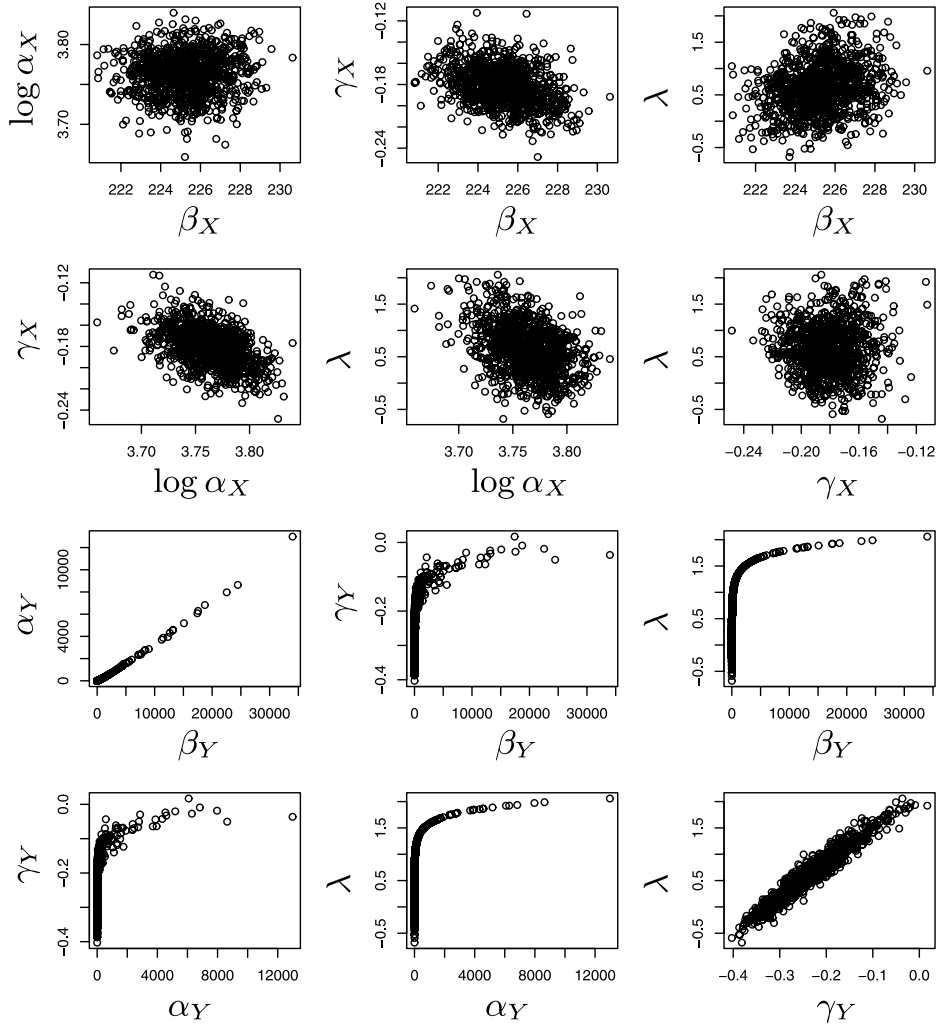}

\caption{Transformed extreme value model example: pairwise
empirical posteriors for the new parameters $\{\beta_X,\log\alpha
_X,\gamma_X,\lambda\}$ (top two rows) and the implied posteriors for
the original parameters $\{\beta_Y,\alpha_Y,\gamma_Y,\lambda\}$
(bottom two rows).}
\label{fig:Ex1Post}
\end{figure}

Figure~\ref{fig:Ex1Post} displays the pairwise empirical posteriors
from the MCMC output. The first two rows exhibit pairs from the new
parameters $\{\beta_X,\log(\alpha_X),\break\gamma_X,\lambda\}$, while
the bottom two rows present the implied posteriors for the original
parameter set $\{\beta_Y,\alpha_Y,\gamma_Y,\lambda\}$. It is clear
from these figures that no meaningful inference could be performed
without the reparameterization.

%f1 ###

%f2 ###

%s4.1.2 ###
\subsubsection{Normal distribution}
\label{sec:NormalDistribution}

The data simulated were 100,000 truncated (at 0) $\mbox{N}(0,1)$
variables, that is, such that $F_X(x)$ = $2\Phi(x)-1, x>0$. As
in the example of Section~\ref{sec:ExtremeValueModel}, three data sets
were obtained from these:
\begin{enumerate}[1.]
\item Block maxima: 1000 block maxima taken over block length 100.
\item 1000 largest data points.
\item All data points above the smallest block maximum. There were 8066
such points.
\end{enumerate}

\begin{figure}[t]

\includegraphics{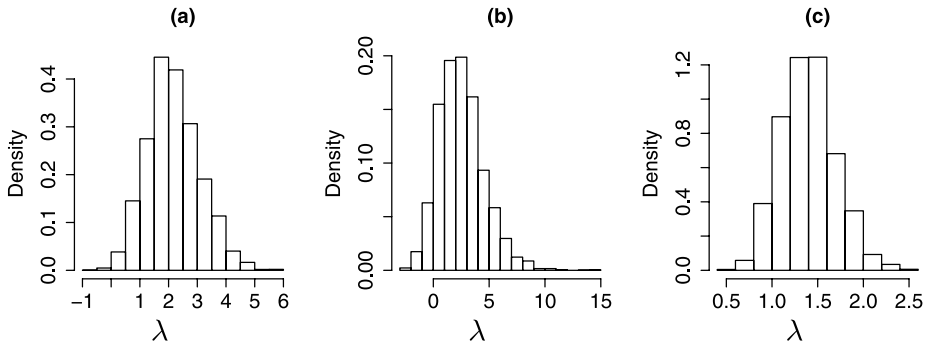}

\caption{Truncated normal example: posteriors for $\lambda$,
for \textup{(a)} data set~1, prior range $[-1,6]$; \textup{(b)}~data set~2,
prior range $[-3,15]$; and \textup{(c)} data set~3, prior range $[0,3]$.}
\label{fig:XiLamProfLambdaPostNormal}
\end{figure}

Figure~\ref{fig:XiLamProfLambdaPostNormal} presents the posteriors for
$\lambda$ in each case. The pattern of information contained on
$\lambda$ from each data set is similar to the previous example, for
the reasons formerly described. In Figure~\ref
{fig:XiLamProfLambdaPostNormal}(a) there is a mode just below $\lambda
=2$, in (c) the peak is around $\lambda=1.5$. These values fit well
with the theory. The location normalizing constant for the truncated
normal distribution is $b_{X,n} \approx(2\log n)^{1/2}-(1/2)\times
(2\log n)^{-1/2}(\log\pi+\log\log n) \approx2.6$ when $n=100$. At
this sub-asymptotic level, the value of $\lambda_n^*$ from~\eqref
{lam}, using the first four leading terms in $h_X/x$ and $h_X'$ is
1.86. For the third data set we are at an even lower asymptotic level.
Here, replacing $b_{X,n}$ in the calculation with the threshold, 1.75,
gives $\lambda_n^* = 1.48$. Both of these agree with the evidence in
the posterior for $\lambda$.

\begin{figure}[t]

\includegraphics{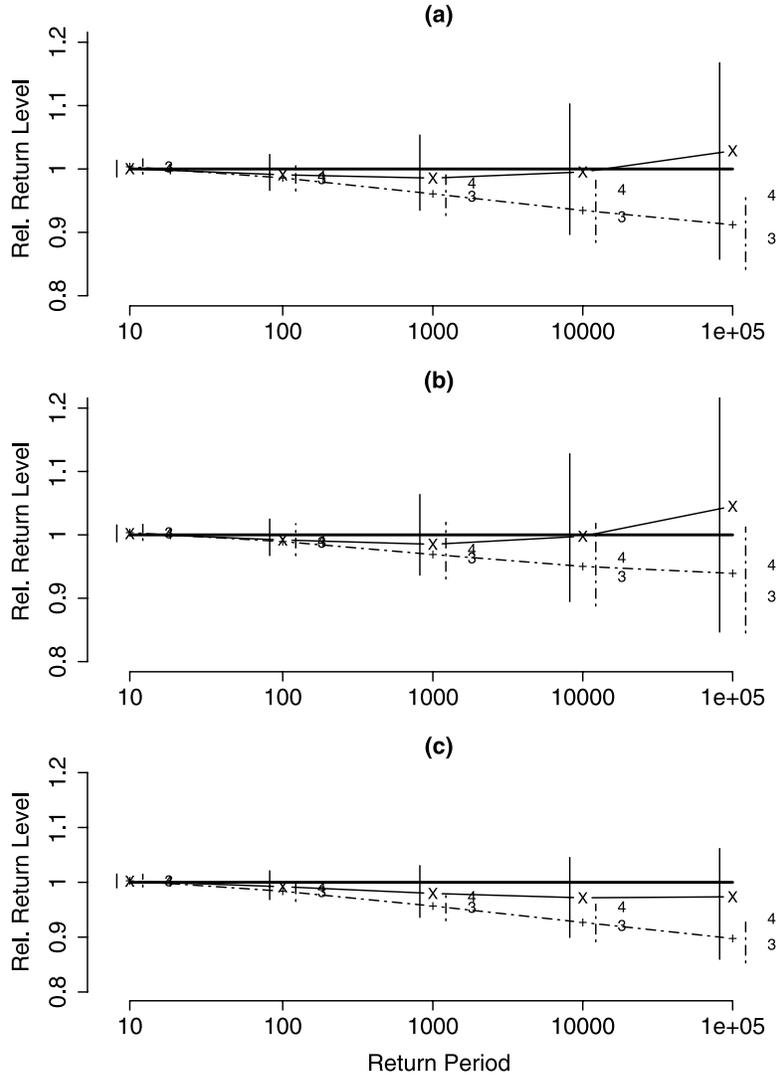}

\caption{Truncated normal example: posterior and posterior
predictive relative return level summaries for \textup{(a)} data set~1,
\textup{(b)} data set~2, \textup{(c)} data set~3. The solid bold line at
1 is the reference point for the true return level based on the
truncated normal cdf; `3', `4' denote the relative posterior median
return levels of the 3 and 4 parameter models respectively;
dashed/solid vertical lines: \textup{3/4} parameter model 95\% credibility interval;
dashed/solid connected lines: \textup{3/4} parameter model posterior
predictive return levels.}
\label{fig:RLNormal}
\end{figure}

Figure~\ref{fig:RLNormal} displays the relative return level summaries
derived from the analysis, with reference to the true return level
curve calculated by solving $\{F_X(x_{1/p})\}^{100} = 1-p$. Posterior
return level summaries are displayed pointwise, while the posterior
predictive distributions are given as curves. In Figures~\ref
{fig:RLNormal}(a) and (c) it can be observed that the 3 parameter
model produces biased estimates of the return levels, the true value
falling far outside the posterior credible interval. In Figure~\ref
{fig:RLNormal}(b) the true value is just covered by the interval.
These results are an indication of the very slow convergence of the
Normal distribution to the extreme value limit. From the posteriors for
$\lambda$ there is certainly evidence that accelerated convergence is
obtained from the 4 parameter model. The bias in return level
estimation compared to the 3 parameter case is reduced, but has not
disappeared. The true values of the return level lie within each of the
credibility intervals for the 4 parameter models. This is in part down
to the faster convergence, although the extra uncertainty involved
plays a role as well.

%f3 ###

%f4 ###

%f5 ###
\begin{figure}

\includegraphics{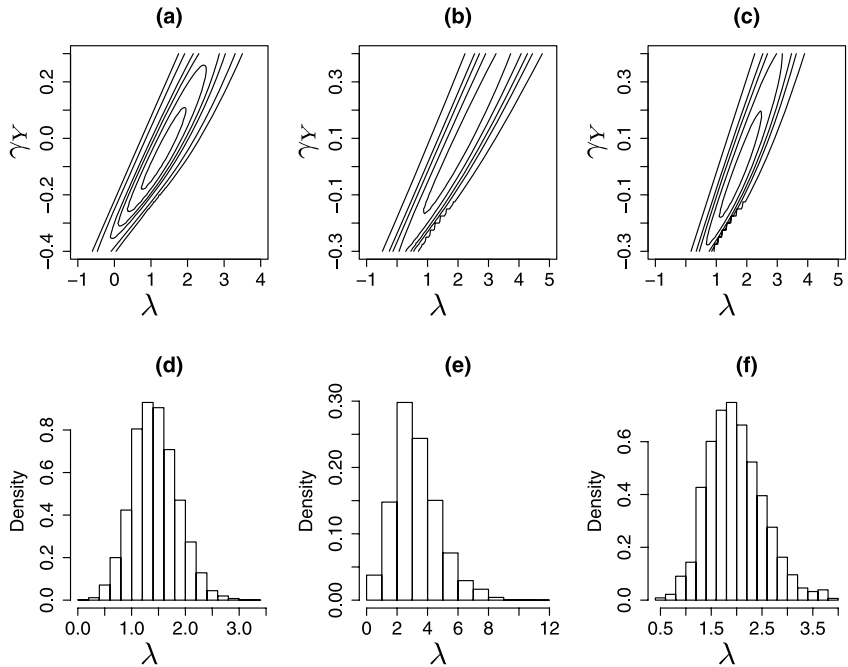}

\caption{$H_s$ data example: \textup{(a)}, \textup{(b)}, \textup{(c)} profile
log-likelihoods for $\{\gamma_Y,\lambda\}$, with contours at levels
of $-$1, $-$3, $-$5, $-$7, $-$12 below the maximum log-likelihood; \textup{(d)}, \textup{(e)}, \textup{(f)}
posteriors for $\lambda$ for analyses \textup{(i)}, \textup{(ii)} and
\textup{(iii)} respectively. Prior ranges for $\lambda$ taken as
\textup{(i)} $[-1,4]$, \textup{(ii)} $[-2,15]$, \textup{(iii)} $[0,5]$.}
\label{fig:LambdaPostNCAll}
\end{figure}

%f6 ###
\begin{figure}[t]

\includegraphics{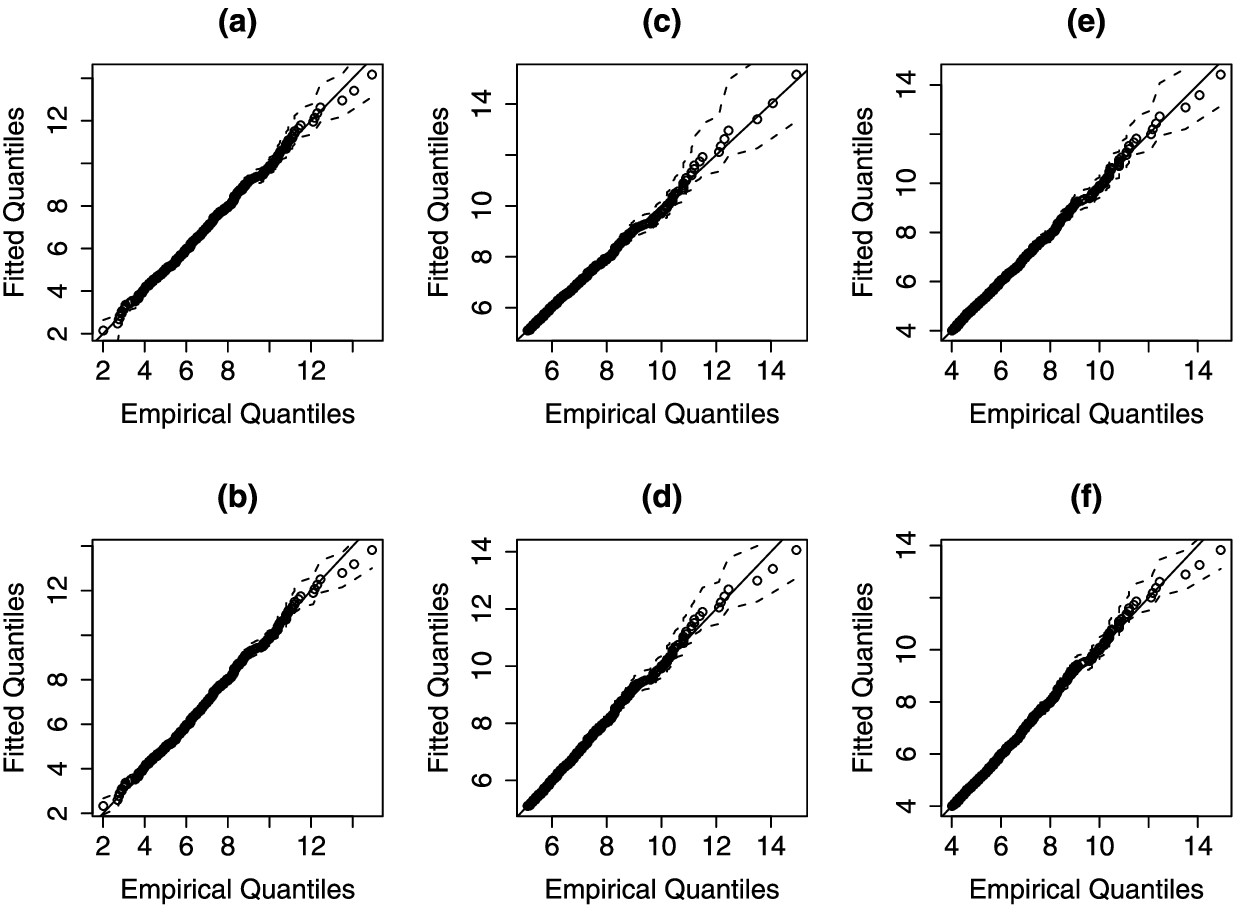}

\caption{$H_s$ data example: QQ plots for \textup{(a)}, \textup{(c)}, \textup{(e)} 4
parameter model, \textup{(b)}, \textup{(d)}, \textup{(f)} 3 parameter model for
analyses \textup{(i)}, \textup{(ii)} and \textup{(iii)} respectively. Dashed lines
represent a 95\% pointwise credible interval, formed from the central
95\% of the posterior distribution for each quantile.}
\label{fig:QQNCAll}
\end{figure}

%s4.2 ###
\subsection{Wave example}
\label{sec:RealDataExample}

The data are measured significant wave heights ($H_s$) for an unnamed
location in the North Sea. There were just over 33 years of
measurements available, with 8 measurements per day recording $H_s$
over continuous 3 hour time periods. Our analysis is restricted to a
single season to ensure approximate stationarity, taking the winter
period (13 weeks beginning on 1 December each year), as this generally
represents the period when almost all extreme events arise. We again
examined the data in three ways:

\begin{enumerate}[(iii)]
\item[(i)] Weekly maxima, corresponding to a block size of $8\times7=56$
observations. There were 433 data points in total.
\item[(ii)] Cluster maxima above an 80\% threshold. Runs method
declustering [\citet{SmithWeiss}] was used, with a separation of 6
consecutive sub-threshold values deemed to define a new cluster. There
were 562 data points.
\item[(iii)] Cluster maxima above an 60\% threshold, using the
same declustering procedure as in (ii). There were 618 data points.
\end{enumerate}
In each case both the usual 3 parameter model and the appropriate
proposed 4 parameter model [GEV for (i), point process for
(ii) and (iii)] were fitted. Our analyses are again based
on 10,000 MCMC samples following a 1000 iteration burn-in period.
Figure~\ref{fig:LambdaPostNCAll} displays the profile likelihoods for
$\{\gamma_Y,\lambda\}$ and the posterior distributions of $\lambda$
in each scenario. As with the simulated data, there is more information
on $\lambda$ for less extreme data, as evidenced by plots (d) and (f)
compared with (e). It is interesting to note that for the 4 parameter
GEV model, the slope $c$ in expression~\eqref{xiy2} was estimated as
0.23, showing how the parameterization~\eqref{xiy2} ties in with the
different shape parameters for $H_s$ ($\hat{\gamma}_X=-0.12$) and
$H_s^2$ ($\hat{\gamma}_X=0.11$) mentioned in Section~\ref
{sec:Introduction}: $0.11 = -0.12 + 0.23\times(2-1)$.

Figure~\ref{fig:QQNCAll} displays QQ plots for each of the fits; here
the `fitted' quantile is defined to be the median of the pointwise
quantile posterior distributions, that is, the median of $x_{1/p}$, for
$x_{1/p}$ given by~\eqref{xp}. Each of the fits appears reasonable,
and considering that $\lambda=1$ is plausible under each of the
posteriors, this is perhaps not too surprising. However, in each case,
there is some evidence that the very upper tail is modeled slightly
better by the 4 parameter model.

\begin{figure}

\includegraphics{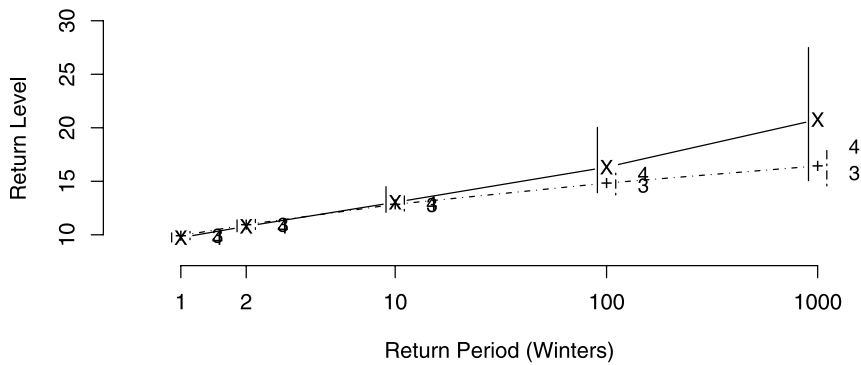}

\caption{$H_s$ data example: posterior and posterior
predictive return level summaries for $H_s$, based on both 3 and 4
parameter models for analysis \textup{(iii)}. Symbols and line types as
in Figure~\protect\ref{fig:RLNormal}.}
\label{fig:RLNC60}
\end{figure}

Posterior summaries of the return levels from analysis~(iii) are
displayed in Figure~\ref{fig:RLNC60}, where increasing disparity of
estimates with lengthening return period can be observed. The
corresponding plots for analyses~(i) and~(ii) have been
omitted for clarity, but show similar general trends with greater
uncertainty for analysis~(ii) and lesser for analysis~(i).
In particular, observe that the medians of the posterior return
level distribution for the 100 and 1000 winter return periods under the
4 parameter model lie into the upper tail of the same distributions
under the 3 parameter model. From the motivating example in
Section~\ref{sec:Introduction} it is clear why these discrepancies
occur: the $H_s$ data were estimated as light-tailed, with a
statistically significant negative shape parameter (taking a 5\%
significance level); the $H_s^2$ data were estimated to be
heavy-tailed, with a statistically significant positive shape
parameter. Such different tail behavior will naturally lead us to
different conclusions. The posteriors for $\lambda$ show that we might
reasonably extrapolate on either scale, among other possibilites; the 4
parameter model combines all such plausible scenarios to build up what
would appear to be a more accurate assessment of the uncertainty
associated with these extrapolations.

%f7 ###

%s5 ###
\section{Discussion}
\label{sec:Discussion}

The paper has presented a parametric method for incorporating the
uncertainty surrounding the scale of extrapolation in extreme value
analysis. Reparameterizations which allow inference under the model
have been derived, justified by the theory of normalizing constants for
the limiting distribution of block maxima. Examples have demonstrated
the ability of the methodology to detect the `true' value of $\lambda$
where one exists, for the case of finite block size/sub-asymptotic
threshold. As either the block size tends to infinity or the threshold
to the upper end point, information on $\lambda$ decreases, since
there is little to be gained from a transformation.

The fact that there may not always be significant information on
$\lambda$ poses the question whether it is always necessary to
incorporate this uncertainty. In Theorem~\ref{th1} we noted that in
the case where $\xi_X\leq0$ with $x^F>0$, the shape parameters $\xi
_{Y,n} \rightarrow\xi_X$ as the data become more extreme, since $\lim
_{x\rightarrow x^F} h_X(x)/x = 0$. In such a case, where all our data
are far into the upper tail, the mean squared error of the 4 parameter
model is likely to exceed that of the 3 parameter case. An
ill-determined posterior for $\lambda$ may be one indication that
utilization of this method adds an unnecessary degree of uncertainty.
If the variance of the posterior seems unacceptably large, then the
suggestion would be that the data do not contain information on
$\lambda,$ in which case the practitioner may consider not using this
method.

At the other end of the scale, the fact that suitable values of
$\lambda$ may accelerate convergence offers the potential for
incorporating more data through lowering of the threshold or
contracting of block length. Although we have not specifically explored
this here, examples such as the normal data example given in
Section~\ref{sec:Examples} demonstrate how this could be worthwhile.
Because QQ plots such as those in Figure~\ref{fig:QQNCAll} are easily
obtained under both 3 and 4 parameter models, the modeler should be
able to determine if there is value in doing this.

A natural question that arises is whether to consider fixing $\lambda$
if there is strong evidence for a particular value in the posterior. As
outlined in Section~\ref{sec:TheoryAndMethods}, there are cases where
a specific value will accelerate convergence, thus, one could assume
that the modal value is a suitable one to take. However, the reason
that we have a full posterior distribution is that there is genuine
uncertainty in this value. Arguably, therefore we mitigate against our
errors by keeping this uncertainty. This seems unsatisfying in a world
where we value precision in our estimates, but if uncertainty genuinely
exists, it should not be masked by the pursuit of false precision.

The Box--Cox class of transformations is suitable only for strictly
positive data. In the event that interest lies in a data set for which
this is not the case, a location shift prior to analysis would be
necessary. One might also in such a case consider different classes of
transformation. \citet{CormannReiss}, for example, consider
exponential transforms. From our proof in the \hyperref[sec:Appendix]{Appendix}
it is simple to derive reparameterizations for any monotonic
transformation, thus, one could exploit this theory in other contexts.

\begin{appendix}
\section*{\texorpdfstring{Appendix: Proof of Theorem
\protect\lowercase{\ref{th1}}}{Appendix: Proof of Theorem 1}}
\label{sec:Appendix}

Denote the transformation $y(x) = \{x^\lambda-1\}/\lambda$ and the
inverse transformation $x(y) = \{\lambda y +1\}^{1/\lambda}$. The
distribution function $F_Y$ is given by
\begin{eqnarray*}
F_Y(y) = \mathsf{P}(Y\leq y) = \mathsf{P}(X\leq\{\lambda y +1\}
^{1/\lambda})
= F_X(\{\lambda y +1\}^{1/\lambda}) = F_X(x(y)).
\end{eqnarray*}
Therefore, solving $F_Y(b_{Y,n}) = 1-1/n$ for $b_{Y,n}$ yields
\begin{eqnarray*}\label{bny1}
F_X(\{\lambda b_{Y,n}+1\}^{1/\lambda}) &=& 1-1/n, \nonumber
\\
\{\lambda b_{Y,n}+1\}^{1/\lambda} &=& F_X^{-1}(1-1/n) = b_{X,n},\nonumber
\\
b_{Y,n} &=& \frac{b_{X,n}^\lambda-1}{\lambda}.
\end{eqnarray*}

Denote the Jacobian of the transformation and inverse transformation by
\begin{eqnarray*}
J_X(x) := \biggl|\frac{dy}{dx}\biggr| = x^{\lambda-1},\qquad  J_Y(y) := \biggl|\frac
{dx}{dy}\biggr| = \{\lambda y +1\}^{1/\lambda-1}.
\end{eqnarray*}
These are linked by $J_Y(y) = \{J_X(x(y))\}^{-1}$. The reciprocal
hazard function $h_Y$ is
\begin{eqnarray*}
h_Y(y) = \frac{1-F_Y(y)}{f_Y(y)} = \frac
{1-F_X(x(y))}{f_X(x(y))J_Y(y)} = \frac{h_X(x(y))}{J_Y(y)},
\end{eqnarray*}
which gives
\begin{eqnarray*}%\label{any1}
a_{Y,n} = h_Y(b_{Y,n}) = \frac{h_X(\{\lambda b_{Y,n} +1\}^{1/\lambda
})}{\{\lambda b_{Y,n}
+1\}^{1/\lambda-1}} = \frac{h_X(b_{X,n})}{(b_{X,n})^{1-\lambda}} =
a_{X,n}(b_{X,n})^{\lambda-1}.
\end{eqnarray*}

To obtain an expression for the shape parameter, we require the
derivative of the reciprocal hazard function for $Y$,
\begin{eqnarray*}
h_Y'(y) &=& \frac{d}{dy}\biggl\{\frac{h_X(x(y))}{J_Y(y)}\biggr\}
\\
&=&{\biggl(J_Y(y)\frac{d}{dy}h_X(x(y))-h_X(x(y))\frac{d}{dy}J_Y(y)\biggr)\big/J_Y(y)^2}.
\end{eqnarray*}

By the chain rule,
\begin{eqnarray*}
\frac{d}{dy}h_X(x(y)) =J_Y(y)h'_X(x(y))
\end{eqnarray*}
and
\begin{eqnarray*}
J'_Y(y) = J_Y(y)\frac{d}{dx}\frac{1}{J_X(x(y))} = -\frac
{J'_X(x(y))}{J_X(x(y))^3}.
\end{eqnarray*}
Thus,
\begin{eqnarray*}
h'_Y(y) &=& \frac{J_Y(y)^2 h'_X(x(y))}{J_Y(y)^2} -\frac
{h_X(x(y))J'_Y(y)}{J_Y(y)^2}
\\
&=& h'_X(x(y)) + h_X(x(y))\frac{J'_X(x(y))}{J_X(x(y))}.
\end{eqnarray*}

Substituting in $J_X(x) = x^{\lambda-1}$, $J'_X(x) = (\lambda
-1)x^{\lambda-2}$ results in
\begin{eqnarray*}
h'_Y(y(x)) =h'_X(x) +\frac{h_X(x)}{x}(\lambda-1).%\label{hprime}
\end{eqnarray*}
Substituting in $x=b_{X,n}$ gives~\eqref{xiyn}; taking the limit as
$x\rightarrow x^F$ gives~\eqref{xiy1}.

For the final statement, $\xi_X = \lim_{x\rightarrow x^F} h'_X(x)
\leq0$ implies that\break $\lim_{x\rightarrow x^F}h_X(x)/x = 0$.
\end{appendix}

\section*{Acknowledgments}
The authors acknowledge discussions with Kevin\break Ewans, of Shell
International Exploration and Production, concerning the wave data
analyzed in Section~\ref{sec:Examples}. We further thank the referees
for suggestions which improved the paper.

\printaddresses

\end{document}